\documentclass[a4paper,11pt]{article}
\usepackage{jheppub}
\usepackage{eurosym}
\usepackage{amssymb}
\usepackage{amsfonts,cite}
\usepackage{amsmath}
\usepackage{graphicx}
\usepackage{multirow}
\usepackage{xcolor}
\usepackage{epsfig}
\usepackage{fancyhdr}

\DeclareMathOperator\arctanh{arctanh}

\newbox\mybox

\newcommand\fverb{\setbox\mybox=\hbox\bgroup\verb}
\newcommand\fverbdo{\egroup\medskip\noindent\fbox{\unhbox\mybox}\ }
\newcommand\fverbit{\egroup\item[\fbox{\unhbox\mybox}]}

\abstract{We address the long-standing ``ghost problem" in higher time-derivative theories (HTDTs), where quantisation typically yields sectors with either unbounded spectra or non-normalisable eigenstates; both rendering the theory unphysical. We propose a novel method that preserves the bounded nature of the spectrum in one particular sector while restoring normalisability by employing a non-unitary similarity transformation. Inspired by techniques from pseudo/quasi-Hermitian PT-symmetric quantum mechanics, we construct a non-unitary map between two Hermitian Hamiltonians, converting ghostly sectors into physically viable ones. We demonstrate the feasibility of this approach using a concrete HTDT model, related to the Pais-Uhlenbeck oscillator, and show that the transformed system admits normalisable eigenstates and a spectrum bounded from below. This framework offers a consistent re-interpretation of HTDTs and extends the toolbox for constructing ghost-free quantum models.    }    

\title{\bfseries Ghost-Free Quantisation of Higher Time-Derivative Theories via Non-Unitary Similarity Transformations}

\author[a]{Andreas Fring,}
\author[b]{Takano Taira,}
\author[a]{Bethan Turner}

\affiliation[a]{Department of Mathematics, City St George's, University of London, Northampton Square, \\ London EC1V 0HB, UK}
\affiliation[b]{Department of Physics, Kyushu University
	744 Motooka, Nishi-ku, Fukuoka City, 819-0395, Japan}

\emailAdd{a.fring@city.ac.uk}
\emailAdd{taira.takano.292@m.kyushu-u.ac.jp}
\emailAdd{bethan.turner.2@city.ac.uk}

\keywords{Higher Time-Derivative Theories, quantisation}

\begin{document}
	\maketitle
	
	\pagestyle{fancy}
	\fancyhead{} % clear all header fields
	\fancyhead[LE,RO]{\small\itshape Ghost-Free Quantisation of HTDT via Non-Unitary Similarity Transformations} 
	
	\renewcommand{\headrulewidth}{0.4pt}

\section{Introduction}	
Higher time derivative theories (HTDTs) are attractive for several reasons, the main one being that they are renormalisable \cite{pais1950field,stelle77ren,grav1,grav2,grav3,modesto16super} and therefore offer a potential framework for unifying gravity and quantum mechanics  \cite{Hawking}. However, they suffer from significant deficiencies that have so far prevented them from being accepted as fully viable physical theories. When quantising such systems, one inevitably encounters at least two sectors with qualitatively different behaviour. In one sector, the eigenstates are normalisable, but the corresponding spectra are unbounded from below. In the other, the eigenstates are non-normalisable, yet the spectra are bounded \cite{weldon03quant,Woodard1,fring2024higher,fring2025quant}. Both sectors, therefore, possess features that would render the theory unphysical under standard interpretations of quantum mechanics.

Several proposals have been made to address these deficiencies, with most focusing on the sector that possesses normalisable eigenfunctions, as it allows for concrete quantum mechanical calculations \cite{ghostconst,salvio16quant,fakeons,bender2008no,raidal2017quantisation}. Especially promising are approaches that base their analysis on the quantisation of classical sectors of the theory that possess well-defined finite solutions in phase space \cite{smilga2006ghost,smilga2021benign,deffayet22ghost,deffayet23global,diez2024foundations}. In this approach one seeks alternative Hamiltonian structures that preserve the dynamics but yield bounded spectra when the theory is quantised \cite{bolonek2005ham,dam2006,stephen2008ostro,most2010h,andrzejewski2014ham,and2014conflett,andrzejewski2014conf,elbistan2023various}. This often entails identifying hidden symmetries or exploiting the systems’ Bi-Hamiltonian structure to develop equivalent formulations with more desirable properties \cite{FFT}.

Here, we propose a different scheme that allows to make sense of the sector with non-normalisable eigenstates and bounded spectra by employing a non-unitary equivalence map to a system with normalisable eigenstates. The situation is akin to that of pseudo/quasi-Hermitian ${\cal PT}$-symmetric quantum mechanics \cite{Bender:1998ke,Alirev,PTbook,fring2023intro}. In that context the usual starting point is often a non-Hermitian Hamiltonian that has a well-defined discrete spectrum of real eigenvalues, but ill-defined inner products of the corresponding wavefunctions. The resolution that leads to a self-consistent well-defined theory is to non-unitarily map the non-Hermitian Hamiltonian to a Hermitian one and employ the map used to define a new metric for the inner product, thus making perfect sense of the non-Hermitian system. Unlike as in the complex scaling approach \cite{bender2008no} the system is considered on the real axis and not rotated to the complex plane.

We modify this approach here and map a Hermitian HTDT or ghostly Hamiltonian $h_0=h_0^\dagger$ with eigenvalue equation $h_0 \phi_0 =E \phi_0$, to another Hermitian Hamiltonian $h_f=h_f^\dagger$ with eigenvalue equation $h_f \phi_f =E \phi_f$ by means of
\begin{equation}
	\eta h_0 \eta^{-1} = h_f, \qquad  \phi_f = \eta \phi_0 ,   \label{simtrans}
\end{equation}  
where $\eta$ is a non-unitary map. We are focusing especially on the sector of solutions for $h_0$ that has a bounded real spectrum, but non-normalisable wave functions $\phi_0$. Evidently, being related by a similarity transformation, the spectra of $h_0$ and $h_f$ are identical. When defining the inner product as  
\begin{equation}
	  \langle  \phi'_f \vert \phi_f      \rangle   =  \langle  \phi'_0 \vert  \rho  \phi_0      \rangle = : \langle  \phi'_0 \vert   \phi_0      \rangle_\rho    \label{newinnprod}
\end{equation}  
with new metric $\rho = \eta^\dagger \eta $ the two spectrally equivalent theories still have  a bounded real spectrum and potentially normalisable wavefunctions in some part of the parameter space. Provided this section of the parameter space is not empty, the  HTDT or ghostly system based on the Hermitian Hamiltonian $h_0$ is now well-defined, i.e. the bound state spectra are bounded from below {\em and} the eigenstates are normalisable. 

We now verify for a concrete example that such a system can be constructed.

\section{From Hermitian to non-Hermitian to Hermitian Hamiltonians}	

\subsection{The $h_0$ eigensystem}

Our sample model is a ghostly two-dimensional Hermitian Hamiltonian system that was recently fully quantised in \cite{fring2025quant}. Moreover, in \cite{FFT} it was established that in some parts of the parameter space the model can be mapped to a prototype HTDT, the Pais-Uhlenbeck model \cite{pais1950field}, so that our findings directly apply to those type of systems. The Hamiltonian is given by 
\begin{equation}
	h_0  =    p_x^2 - p_y^2   + \nu^2 x^2+  \Omega  y^2  + g x y, \qquad \nu,\Omega,g   \in \mathbb{R} , \label{Hghost}
\end{equation}
with the negative sign in front of the $p_y$-term indicating its ghostly nature.
The Hamiltonian is fully $\cal{PT}$-symmetric under $\cal{PT}:$ $p_x \rightarrow p_x$, $p_y \rightarrow p_y$, $x \rightarrow -x$, $y \rightarrow -y$ and $i\rightarrow - i$ and is therefore set up for the use of techniques developed in context of pseudo/quasi-Hermitian ${\cal PT}$-symmetric quantum mechanics \cite{Bender:1998ke,Alirev,PTbook,fring2023intro}. The entire eigensystem of $h_0$, for the eigenvalue equation $ 	h_0  \phi = E \phi $, was constructed in \cite{fring2025quant}. 
The energy spectrum was found to be
\begin{eqnarray}
	E_{Nn}^\pm &=& (N+1) (\alpha -\beta )\pm (2-2 n+N)\sqrt{ (\alpha +\beta )^2-4 \gamma ^2}, \quad n=1, \ldots , \left\lfloor \frac{N}{2}\right\rfloor , \label{eab1} \\
	E_{NN} &=&  (1+N) (\alpha -\beta ), \qquad     (N+1 \bmod 2), \label{eab2}
\end{eqnarray}
where the parameters $\alpha,\beta, \gamma$ stem from the Ansatz for the wavefunctions and are related to the model parameters $\nu,\Omega,g$ as 
\begin{equation}
	\alpha_\epsilon^\eta = \frac{2 \nu ^2  + \sigma_\epsilon }{ \Sigma_\epsilon^\eta}, \quad
	\beta_\epsilon^\eta = \frac{2 \Omega - \sigma_\epsilon}{\Sigma_\epsilon^\eta}, \quad
	\gamma_\epsilon^\eta = \frac{-g}{ \Sigma_\epsilon^\eta }, \,\, \Sigma_\epsilon^\eta = 2 \eta \sqrt{ \nu ^2-\Omega+ \sigma_\epsilon },
	\,\, \sigma_\epsilon =  \epsilon \sqrt{g^2-4 \nu ^2 \Omega } ,  \label{abcd}
\end{equation}
constituting four different branches of the theory labelled by all combinations of $\epsilon = \pm 1$, $\eta=\pm 1$. In particular, the sector $\epsilon=\eta=1$ is of interest to us here as it possesses a spectrum that is bounded from below, see also figure 2 in \cite{fring2025quant}. 

The eigenfunctions were found to be of Gaussian form multiplied by polynomials in $x,y$. The normalisability is entirely captured by the ground state, which passes on its behaviour to all excited states. Thus, ensuring the normalisability of the ground state is sufficient to guarantee the normalisability of all excited states. In \cite{fring2025quant} we identified this state and the corresponding eigenvalue equation as
\begin{equation}
	\phi_0(x,y) = c_0 e^{-\frac{\alpha  x^2}{2}-\frac{\beta  y^2}{2} +\gamma  x y   } , \qquad   \text{with} \quad  h_0 	\phi_0(x,y) =(\alpha -\beta)  	\phi_0(x,y) ,\label{wansatz}
\end{equation}
and $c_0$ denoting some overall normalisation constant.  It turned out that the two sectors build on $\epsilon = -1$ possess wavefunctions that are normalisable, but have spectra that are not bounded from below. In turn the two $\epsilon = 1$ sectors have non-normalisable wavefunctions, but bound spectra from below, when $\eta =1$, and above, when $\eta =-1$. Concretely, the normalisability condition 
\begin{equation}
	\alpha > 0, \qquad \beta > 0, \qquad   \alpha \beta-  \gamma^2>0,    \label{normcond}
\end{equation} 
can only be satisfied for the two solutions with $\epsilon = -1$. We will now describe how to make sense of the $\epsilon = 1$ sectors.

\subsection{From $h_0$ to $H_1$}

Our aim is to find a suitable similarity transformation as introduced in (\ref{simtrans}), or in fact similarity transformations. We do not expect these transformations to be unique, for the same well-known reasons as in the context of pseudo/quasi-Hermitian ${\cal PT}$-symmetric quantum mechanics \cite{Urubu}. Suitable candidates of operators to be used are
\begin{equation}
	\eta_0 = \exp\left(  -\frac{\delta}{2} x^2 -\frac{\lambda}{2} y^2     \right), \quad
	\eta_1 = \exp\left(  \frac{\kappa}{2} p_x^2 +\frac{\xi}{2} p_y^2     \right), \quad
	\eta_2 = \exp\left( \mu p_x p_y + \tau x y   \right),
\end{equation} 
with $\delta, \lambda, \kappa, \xi, \mu, \tau \in \mathbb{R}   $. We compute their adjoint actions on the canonical variables $x,y,p_x,p_y$ by means of the Baker-Campbell-Hausdorff formula\footnote{See e.g. chapter 5 in \cite{hall2015lie}: For two operators $A$ and $B$ we have
	\[
	e^A B e^{-A} = B + [A, B] + \frac{1}{2!}[A, [A, B]] + \frac{1}{3!}[A, [A, [A, B]]] + \cdots
	\]}, as it is usually referred to in the physics literature, 
\begin{eqnarray}
	\eta_0 x \eta_0^{-1} &=&x, \quad \,\,\, \,\, \eta_0 y \eta_0^{-1} =y, \quad
	\eta_0 p_x \eta_0^{-1} =p_x - i \delta x,  \quad \eta_0 p_y \eta_0^{-1} =p_x - i \lambda y,  \label{eta0} \\
	\eta_1 p_x \eta_1^{-1} &=&p_x, \quad  \eta_1 p_y \eta_1^{-1} =p_y, \quad
	\eta_1 x \eta_1^{-1} =x - i \kappa p_x,  \quad \, \, \eta_1 y \eta_1^{-1} =y - i \xi p_y, \label{eta1} \\
	\eta_2 x \eta_2^{-1} &=&  \cosh ( \theta ) x - i \frac{\mu}{\theta} \sinh( \theta) p_y, \qquad 
		\eta_2 y \eta_2^{-1} =  \cosh ( \theta ) y - i \frac{\mu}{\theta} \sinh( \theta) p_x, \qquad \label{eta2} \\
		\eta_2 p_x \eta_2^{-1} &=&  \cosh ( \theta ) p_x + i \frac{\tau}{\theta} \sinh( \theta) y, \qquad 
		\eta_2 p_y \eta_2^{-1} =  \cosh ( \theta ) p_y + i \frac{\tau}{\theta} \sinh( \theta) x, \label{eta22}
\end{eqnarray} 
with $\theta := \sqrt{\mu \tau}$. Acting adjointly with these operators on $h_0$  in different orders will yield different types of Hamiltonians. Our ultimate aim is to generate a target Hamiltonian that is Hermitian. Starting with $\eta_0$, using (\ref{eta0}), we can only map $h_0$ to a non-Hermitian Hamiltonian\footnote{Our convention is here to denote Hermitian Hamiltonians by $h$, non-Hermitian Hamiltonians by$H$, and their corresponding eigenfunctions by $\phi$ and $\psi$, respectively.} by means of a similarity transformation 
\begin{equation}
	H_1= \eta_0 h_0 \eta_0^{-1} =   p_x^2 -  p_y^2 + (\nu^2 - \delta^2) x^2 + (\Omega + \lambda^2) y^2 + g x y
	- i \delta (p_x x + x p_x ) + i \lambda (p_y y + x p_y) .   \label{nonH1}
\end{equation}
We notice that $H_1$ is fully $\cal{PT}$-symmetric, although not a standard system considered in the context of $\cal{PT}$-symmetric quantum mechanics as its $(+,-)$-signature in the kinetic term indicates its ghostly nature. Since $\eta_0$ does not depend on the momenta, the corresponding eigenfunction $\psi_1 $ is trivially obtained from a simple scalar multiplication
\begin{equation}
	\psi_1(x,y) =   \eta_0(x,y) \phi_0(x,y)= c_1 e^{-\frac{(\alpha+ \delta)  x^2}{2}-\frac{(\beta+\lambda)  y^2}{2} +\gamma  x y   },
\end{equation} 
with $c_1$ being a normalisation constant. Tuning the newly available parameters $\delta, \lambda$, this wavefunction is already normalisable in some parameter regime in the sense of (\ref{newinnprod}). However, when including the excited states into the spectrum, we will not end up with a positive definite orthonormal system for the well-known reasons, see  \cite{Bender:1998ke,Alirev,PTbook,fring2023intro}, since $H_1$ is non-Hermitian. Thus, we continue by exploring the action of other additional operators.

\subsection{From $H_1$ to $H_2$/$h_2$}

Continuing in this way we construct a second non-Hermitian Hamiltonian by means of a similarity transformation using $\eta_1$
\begin{eqnarray}
	H_2&=& \eta_1 H_1\eta_1^{-1} =  \frac{\nu^2}{\nu^2 - \delta^2}  p_x^2 - \frac{\Omega}{\lambda^2 + \Omega}  p_y^2 + (\nu^2 - \delta^2) x^2 + (\Omega + \lambda^2) y^2 + g x y \\
	  && + \frac{g \delta \lambda}{ (\nu^2 - \delta^2)(\lambda^2 + \Omega)} p_x p_y
	+ i \frac{ g \delta}{\nu^2 - \delta^2} p_x y  + i \frac{g \lambda}{\lambda^2 + \Omega}  p_y x , \notag
\end{eqnarray}
where we made the choices  $ \xi= \lambda/(\lambda^2 + \Omega) $ and $ \kappa= \delta /(\delta^2 - \nu^2) $ to eliminate non-Hermitian terms proportional to $ i p_x x$ and $ i p_y y$. We notice that also $H_2$ is $\cal{PT}$-symmetric and moreover, when $g \rightarrow 0$ it becomes Hermitian and simply a sum of two harmonic oscillators in a certain parameter regime. Thus, in this case we have already achieved our goal and mapped $h_0$ to another Hermitian Hamiltonian $h_f=h_2$.

Because $\eta_1$ depends on the momenta, we compute its action on the wavefunction in momentum space. Noting that in Fourier space derivatives become multiplications, we compute  
\begin{equation}
	\psi_2(x,y) =  \eta_1(p_x,p_y) 	\psi_1(x,y) = {\cal F}^{-1}_{xy} \left[ \eta_1(p_x,p_y)  \hat{\psi}_1(p_x,p_y)     \right] (x,y),
\end{equation}
where $\hat{\psi}_1(p_x,p_y)  $ is the Fourier transform of $\psi_1(x,y)$ and $ {\cal F}^{-1}_{xy}$ denotes the inverse Fourier transform. Carrying out the transformations we obtain
\begin{equation}
	\psi_2(x,y) = c_2 e^{-\frac{ \hat{\alpha}  x^2}{2}-\frac{ \hat{\beta}  y^2}{2} +\hat{\gamma}  x y   },
\end{equation}
with normalisation constant $c_2$ and
\begin{eqnarray}
	\hat{\alpha} &=&  \frac{ \left[(\alpha +\delta ) (\beta  \lambda -\Omega )-\gamma ^2 \lambda \right] \left(\delta ^2-\nu ^2\right) }{\left(\alpha  \delta +\nu ^2\right)
		(\Omega -\beta  \lambda )+\gamma ^2 \delta  \lambda }, \quad \label{alphh1} \\
	\hat{\beta} &=& \frac{\left[(\beta +\lambda ) \left(\alpha  \delta +\nu ^2\right)-\gamma ^2 \delta \right]  \left(\lambda ^2+\Omega \right) }{\left(\alpha  \delta +\nu
		^2\right) (\Omega -\beta  \lambda )+\gamma ^2 \delta  \lambda }, \\
	\hat{\gamma} &=& - \frac{  \gamma   \left(\delta ^2-\nu ^2\right)  \left(\lambda ^2+\Omega \right)     }{\left(\alpha  \delta +\nu
		^2\right) (\Omega -\beta  \lambda )+\gamma ^2 \delta  \lambda } .  \label{alphh3} 
\end{eqnarray}

\begin{figure}[h]
	\centering         
	\begin{minipage}[b]{0.54\textwidth}     
		\includegraphics[width=\textwidth]{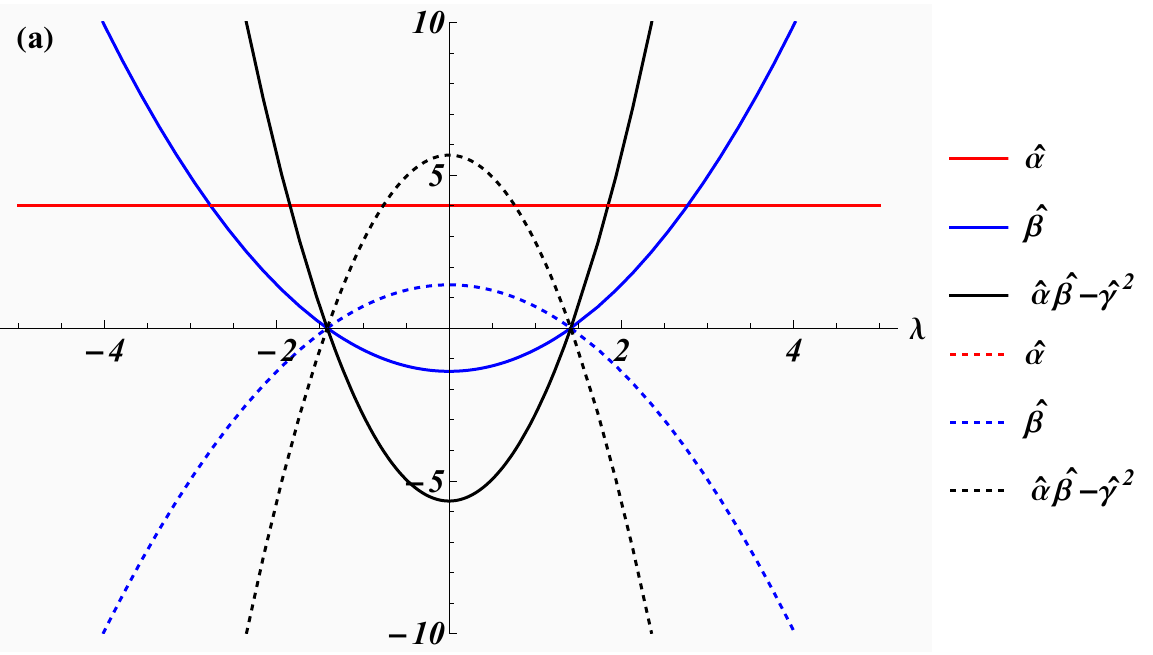}
	\end{minipage}   
	\begin{minipage}[b]{0.44\textwidth}           
		\includegraphics[width=\textwidth]{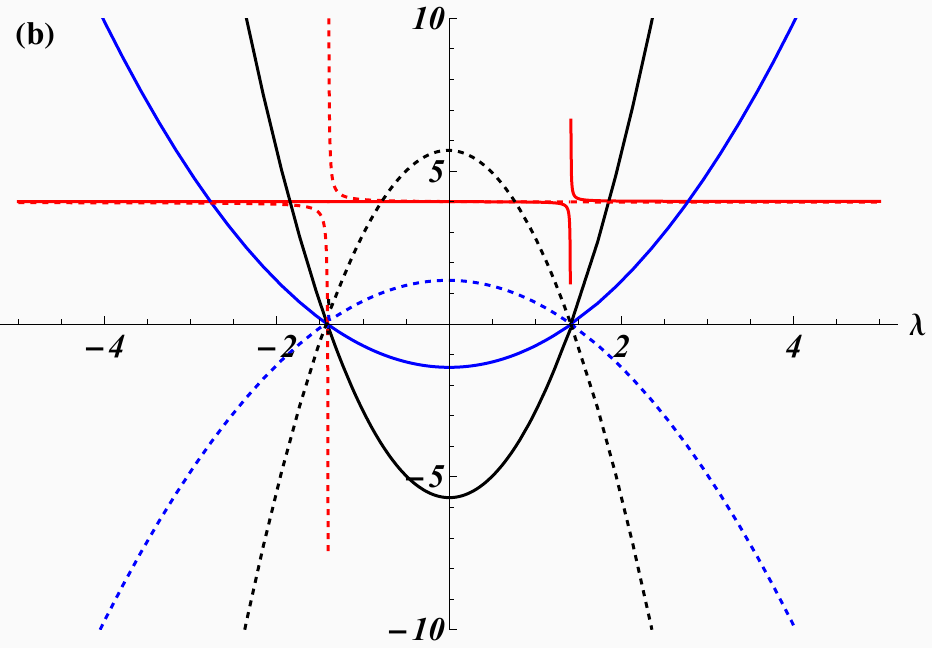}
	\end{minipage}    
	\caption{Normalisability condition (\ref{normcond}) for the parameters $\hat{\alpha}$, $\hat{\beta}$, $\hat{\gamma}$ in (\ref{alphh1})- (\ref{alphh3}) with $\delta=0$, $\nu=4$, $\Omega=-2$ and $g=0$ in panel (a), $g=1$ in panel (b). Solid and dashed lines correspond to the cases $\epsilon=1$, $\eta=1$ and $\epsilon=-1$, $\eta=1$, respectively.   }
	\label{paraxxx}
\end{figure}

We kept the same choices for $\xi$ and $\kappa$ as in the derivation of $H_2$, and verified that the eigenvalue equation $H_2	\psi_2 =(\alpha- \beta) \phi_2 $ is indeed satisfied. For these parameter choices, the normalisability condition (\ref{normcond}) can be fulfilled in the $\epsilon=\eta=1$ case where the spectrum is bounded from below. In figure \ref{paraxxx}, the three relevant quantities in (\ref{normcond}) are plotted such that the wavefunction is normalisable when all of them are positive.

As $\delta \rightarrow 0$ and $\lambda \rightarrow 0$, we have $\hat{\alpha} \rightarrow \alpha$, $\hat{\beta} \rightarrow \beta$, and $\hat{\gamma} \rightarrow \gamma$, thus recovering the original behaviour. In this limit, the wavefunction is normalisable for $\epsilon = -1$, $\eta = 1$, and non-normalisable for $\epsilon = 1$, $\eta = 1$, as we can verify in figure \ref{paraxxx}. Upon introducing the parameter $\lambda$, we observe that these features persist in the regime $\vert \lambda \vert < \sqrt{2}$. In the complementary region, $\vert \lambda \vert> \sqrt{2}$, the behaviour is reversed. Thus, in this region the model has become well-defined, possessing a discrete spectrum bounded from below with each eigenstate associated to a normalisable eigenfunction.

Note that in the case of vanishing coupling ($g = 0$), the Hamiltonian $H_2$ corresponds to the difference of two harmonic oscillators when $\vert \lambda \vert < \sqrt{2}$, and to the sum of two harmonic oscillators in the complementary regime. Finally, observe that the parameter $\delta$ has not yet been used; it will merely distort the parameter regime without altering the qualitative behaviour.

\subsection{From $H_2$ to $h_3$}

Next we investigate whether we can also attain a Hermitian target Hamiltonian for non-vanishing coupling constant $g \neq 0$, by utilising the operator $\eta_2$ as a suitable map. Indeed, we calculate the Hermitian Hamiltonian
\begin{eqnarray}
	h_3&=& \eta_2 H_2\eta_2^{-1}  \\ 
	&=& \frac{1}{2} (b_1+b_2^+) p_x^2 + \frac{1}{2} (b_1- b_2^+) p_y^2 +  
	\frac{(\delta^2 - \nu^2)^2 }{2 \delta^2}  \left[ (b_1+ b_2^-) x^2  + (b_1- b_2^-) y^2 \right]\\
	&& +g \frac{\delta ^2 }{\left(\delta ^2-\nu ^2\right)^2} p_x p_y+ g x y    , \notag
\end{eqnarray}
with
\begin{equation}
	b_1 = -\frac{\delta  (\delta +\lambda )}{\delta ^2-\nu ^2}, \qquad  b_2^\pm = \frac{\delta  \lambda -\nu ^2 }{\delta
		^2-\nu ^2}  \sqrt{1- \Theta^2  } \pm 1, \quad \Theta := \frac{g \delta^2 }{\left(\delta ^2-\nu ^2\right) \left(\nu ^2-\delta  \lambda \right)}
\end{equation}
and the constraints
\begin{equation}
	\tau = \frac{\delta ^2-\nu ^2 }{2
		\delta }  \arctanh \Theta , \quad   \mu =  \frac{\delta }{2( \delta^2 - \nu^2   ) }  \arctanh  \Theta, \quad 	\Omega = \frac{ \lambda \left[ \nu^2 - \delta(\delta + \lambda)      \right] }{\delta},
		\quad \left\vert   \Theta \right\vert <1.   \label{const123}
\end{equation}
Without presenting the details here, it is noteworthy that when compared to $h_0$, the transformed Hamiltonian $h_3$ has lost its ghostly nature and possesses regions in parameter space where it is positive definite. 

Crucially, we still need to identify the corresponding eigenfunctions and check their normalisability properties. The ground state is evidently $	\phi_3(x,y) =  \eta_2(x,y,p_x,p_y) 	\psi_2(x,y) $, but since the argument of the exponential function $\eta_2$ depends on a sum of non-commuting operators, we must first Gauss decompose it. Noting that $\eta_2$ can be expressed in terms of rasing and lowering su(2)-generator with $S_- =p_x p_y$, $S_+ = x y$,  $S_z =i/2 (p_x x + y p_y)$ satisfying $[S_z, S_\pm] = \pm S_\pm$, $[S_+, S_-] = 2 S_z$, we make use of  the factorisation
\begin{equation}
\eta_2 = e^{\mu S_- + \tau S_+} =\eta_+ \eta_z \eta_- =e^{\zeta_+ S_+} e^{\ln \zeta_z  S_z } e^{\zeta_- S_-},
\end{equation}
where
\begin{equation}
	\zeta_z := \frac{1}{\cosh^2 \sqrt{\mu \tau} }, \quad \zeta_+ :=\frac{\tau}{ \sqrt{\mu \tau}} \tanh \sqrt{\mu \tau}, \quad \zeta_- :=\frac{\mu}{ \sqrt{\mu \tau}} \tanh \sqrt{\mu \tau},
\end{equation}
see for instance \cite{klimov2009group}. We compute the adjoint action of each factor to 
\begin{eqnarray}
\eta_- x \eta_-^{-1}  &=& x -i \zeta_- p_y      , \quad \eta_- y \eta_-^{-1} = y- i \zeta_- p_x       , \quad \eta_- p_x \eta_-^{-1} = p_x
      , \quad \eta_- p_y \eta_-^{-1} =    p_y ,  \qquad    \\
       \eta_z x \eta_z^{-1}  &=& x \sqrt{\zeta_0}     , \quad \eta_z y \eta_z^{-1} =  y \sqrt{\zeta_0}       , \quad \eta_z p_x \eta_z^{-1} = \frac{p_x}{ \sqrt{\zeta_0}}  
       , \quad \eta_z p_y \eta_z^{-1} =    \frac{p_y} {\sqrt{\zeta_0}  }      ,\\
       \eta_+ x \eta_+^{-1}  &=&  x     , \quad \eta_+ y \eta_+^{-1} =  y     , \quad \eta_+ p_x \eta_+^{-1} = p_x + i \zeta_+ y
       , \quad \eta_+ p_y \eta_+^{-1} =  p_y + i \zeta_+ x      .
\end{eqnarray}
The successive adjoint action of $ \eta_-$, $\eta_z$,  $\eta_+$, naturally gives the same as the direct calculation with $\eta_2$ reported in (\ref{eta2}) and (\ref{eta22}). 

\begin{figure}[h]
	\centering         
	\begin{minipage}[b]{0.535\textwidth}     
		\includegraphics[width=\textwidth]{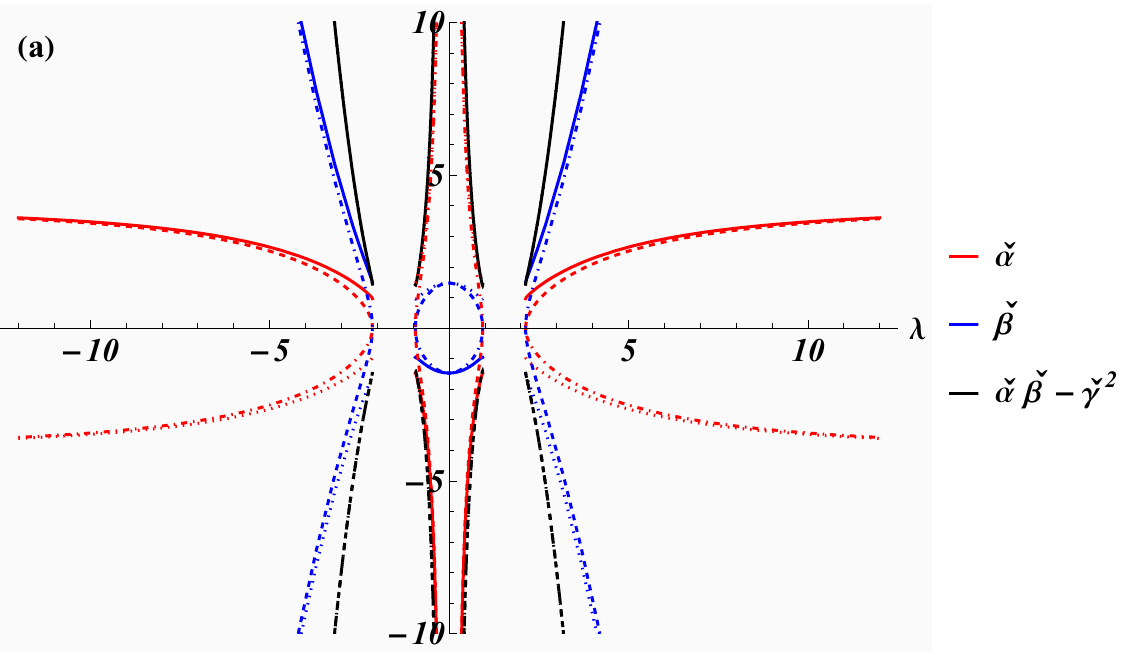}
	\end{minipage}   
	\begin{minipage}[b]{0.445\textwidth}           
		\includegraphics[width=\textwidth]{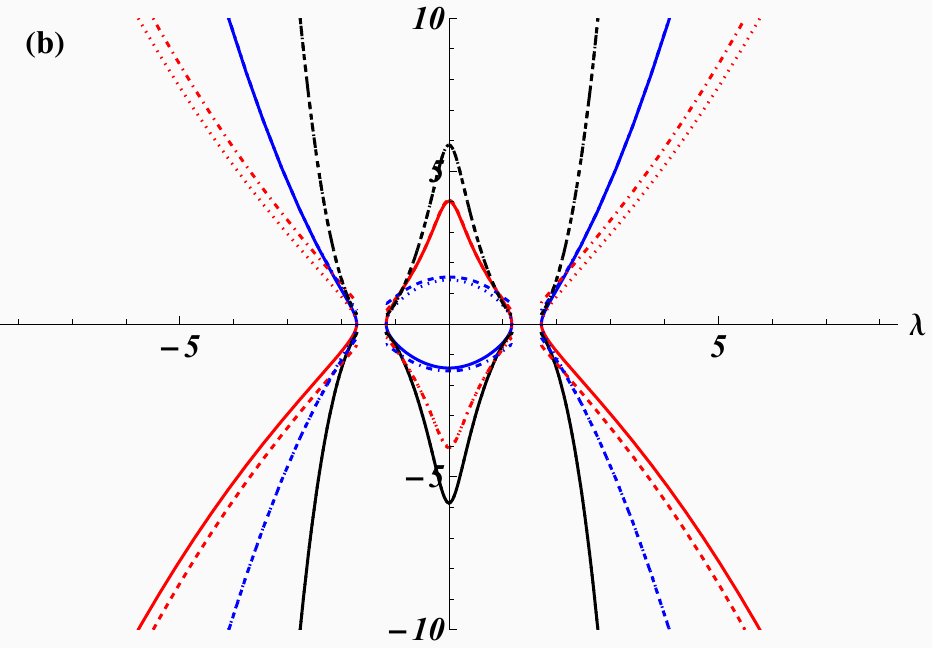}
	\end{minipage}    
	\caption{Normalisability condition (\ref{normcond}) for the parameters $\check{\alpha}$, $\check{\beta}$, and $\check{\gamma}$ in (\ref{checkabc}), with fixed values $\nu = 4$, $\Omega = -2$, and $g = 3$. Panels (a) and (b) correspond to the $\delta_+$ and $\delta_-$ constraints, respectively. The four sectors are distinguished by line styles: solid for $(\epsilon, \eta) = (1, 1)$, dotted for $(1, -1)$, dashed for $(-1, 1)$, and dash-dotted for $(-1, -1)$. }
	\label{normxxx}
\end{figure}

Crucially, we can now act factor by factor on $\psi_2(x,y)$ and calculate
\begin{equation}
	\phi_3(x,y) = e^{\zeta_+ S_+} e^{\ln \zeta_z  S_z } e^{\zeta_- S_-}\psi_2(x,y)= c_3 e^{-\frac{ \check{\alpha}  x^2}{2}-\frac{ \check{\beta}  y^2}{2} +\check{\gamma}  x y   },
\end{equation}
with 
\begin{equation}
	\check{\alpha} = \frac{\hat{\alpha}  \zeta_z}{ \chi}, \quad
	\check{\beta} = \frac{\hat{\beta }\zeta_z  }{ \chi}, \quad
	\check{\gamma} = \frac{\zeta_- (\hat{\gamma}^2 - \hat{\alpha} \hat{\beta} ) +\hat{\gamma}}{\chi} \zeta_z +\zeta_+, \quad
	\chi= \zeta_- ^2 \left(\hat{\gamma} ^2-\hat{\alpha} \hat{\beta} \right)+2 \hat{\gamma}  \zeta_- +1 .
	\label{checkabc}
\end{equation}
At this point it is useful to verify that the eigenvalue equation $h_3 \phi_3 = (\alpha -\beta) \phi_3$ is indeed satisfied. Finally, we examine whether there exists a parameter regime in which the normalisation condition (\ref{normcond}) is satisfied for $\check{\alpha}$, $\check{\beta}$, and $\check{\gamma}$ in the $\epsilon = \eta = 1$ sector. As the parameter space associated with this set of equations is highly complex, we limit ourselves here to a proof of concept by demonstrating that a solution exists for certain parameter choices. The results are shown in figure \ref{normxxx}.

In order to be able to work with fixed values of the original model parameters $\nu,\Omega,g$, we solve the third constraint in (\ref{const123}) for $\delta$, i.e. $\delta_\pm =[ \pm \sqrt{ 4 \lambda^2 \nu^2 +(\lambda^2 + \Omega)^2  } -\lambda^2 -\Omega ]/2 \lambda$, and depict them separately in panels (a) and (b). We find that the final constraint in (\ref{const123}), $\vert \Theta \vert <1$,  identifies two regions where $\check{\alpha}$, $\check{\beta}$, and $\check{\gamma}$ remain real:
R1: $|\lambda| > 2.13322$, and
R2: $|\lambda| < 0.93755$, as seen in figure \ref{normxxx}.

The key result is that for the $\delta_+$ constraint, the normalisability condition (\ref{normcond}) is satisfied for the $\epsilon = \eta = 1$ sector in the R1 region. In contrast, no such solution exists under the $\delta_-$ constraint. Additionally, we confirm that the system remains in the regime where the ground state energy $\alpha - \beta$ is real. For the reasons mentioned above the same logic applied to the excited states.

In summary, we have obtained the isospectral Hermitian Hamiltonian 
\begin{equation}
	h_3 = \eta_2 \eta_1 \eta_0 h_0 \eta_0^{-1} \eta_1^{-1} \eta_2^{-1}.
\end{equation}
The non-normalisable eigenstates $\phi_0$ in the $(1,1)$-sector of $h_0$, defined with respect to the standard metric in the Hilbert space, become normalisable when the metric is changed to $  \rho=  \eta_0^\dagger \eta_1^\dagger \eta_2^\dagger \eta_2 \eta_1 \eta_0$ as specified in (\ref{newinnprod}).

\section{Conclusion}

We have addressed the persistent ghost problem inherent in HTDTs by employing a novel approach based on non-unitary similarity transformations. Rather than discarding the sectors of the theory with non-normalisable eigenstates but a spectrum bounded from below, a sector traditionally considered unphysical, we demonstrated that it can be consistently re-interpreted via a sequence of similarity transformations inspired by pseudo/quasi-Hermitian $\cal{PT}$-symmetric quantum mechanics.

By applying explicitly constructed non-unitary maps to a representative ghostly HTDT Hamiltonian related to the Pais-Uhlenbeck oscillator, we showed how one can systematically transform the system into an equivalent Hermitian model with normalisable wavefunctions and a spectrum still bounded from below. This was achieved without altering the physical energy spectrum of the theory, and while preserving $\cal{PT}$-symmetry throughout. Our results confirm that the ghost problem in HTDTs is not necessarily an insurmountable barrier but can instead be circumvented by re-interpreting the structure of the Hilbert space and inner product.

This framework opens new directions for constructing physically consistent quantum models from otherwise problematic HTDTs and suggests that ghost-free quantisation via non-unitary transformations is a promising addition to the toolkit also for quantum field theory.

Here, we have established a proof of concept, demonstrating the viability of our method in a representative model. Naturally, many open questions remain and will need to be addressed in future investigations to fully uncover the mathematical and physical properties of the transformed systems. As in $\cal{PT}$-symmetric quantum mechanics, the issue of uniqueness persists: different similarity transformations can lead to different final Hamiltonians $h_f$. However, this ambiguity can be resolved by choosing a second observable in addition to the energy \cite{Urubu}.
Extensions to field-theoretic systems can be pursued along similar lines. We believe this approach is broadly applicable to a wide class of HTDTs, if not all.

\medskip

\noindent {\bf Acknowledgments}: BT is supported by a City St George's, University of London Research Fellowship. TT gratefully acknowledges the support of K2-Spring for providing the resources necessary for this research.

\newif\ifabfull\abfulltrue

%%\bibliographystyle{phreport}
%%\bibliography{acompat,Ref}

\end{document}